\documentclass{aa}

\usepackage{graphicx}
\usepackage{amssymb}
\usepackage{amsmath}
\usepackage{natbib}
\bibpunct{(}{)}{;}{a}{}{,}

\def\eg{{\it e.g.\/}\rm,\ }
\def\lsim{\lesssim}
\def\gsim{\gtrsim}

\begin{document}

\title{On the Nature of the  EIS Candidate Clusters: Confirmation of
$z\lsim0.6$ candidates}
\authorrunning{L.F. Olsen et al.}

\author{L.F. Olsen\inst{1} \and C. Benoist\inst{2} \and L. da Costa\inst{3} \and M. Scodeggio\inst{4} \and H.E. J{\o}rgensen\inst{1} \and S. Arnouts\inst{3} \and S. Bardelli\inst{5} \and A. Biviano\inst{6} \and M. Ramella\inst{6} \and E. Zucca\inst{5}}

\institute{
Astronomical Observatory, University of Copenhagen, Juliane Maries Vej 30, DK-2100 Copenhagen, Denmark
\and Observatoire de la C\^{o}te d'Azur, BP 229, 06304 Nice, cedex 4, France 
\and European Southern Observatory, Karl-Schwartzschild-Strasse 2, D-85748 Garching bei M\"{u}nchen, Germany
\and Istituto di Fisica Cosmica, via Bassini 15, I-20133 Milano, Italy
\and Osservatorio Astronomico di Bologna, via Ranzani 1, I-40127 Bologna, Italy
\and Osservatorio Astronomico di Trieste, Via G.B. Tiepolo, 11, I-34131 Trieste, Italy}
\offprints{Lisbeth Fogh Olsen,\\ lisbeth@astro.ku.dk}

\date{Received ; accepted}

\abstract{ We use publicly available $V$-band imaging data from the
wide-angle surveys conducted by the ESO Imaging Survey project (EIS)
to further investigate the nature of the EIS galaxy cluster
candidates.  These were originally identified by applying a
matched-filter algorithm which used positional and photometric data of
the galaxy sample extracted from the $I$-band survey images.  In this
paper, we apply the same technique to the galaxy sample extracted from
$V$-band data and compare the new cluster detections with the original
ones. We find that $\sim75\%$ of the low-redshift cluster candidates
($z\lesssim0.6$) are detected in both passbands and their estimated
redshifts show good agreement with the scatter in the redshift
differences being consistent with the estimated errors of the
method. For the ``robust'' $I$-band detections the matching frequency
approaches $\sim85\%$. We also use the available $(V-I)$ color to
search for the red sequence of early-type galaxies observed in rich
clusters over a broad range of redshifts.  This is done by searching
for a simultaneous overdensity in the three-dimensional
color-projected distance space.  We find significant overdensities for
$\sim75\%$ of the ``robust'' candidates with $z_I\lesssim0.6$. We find
good agreement between the characteristic color associated to the
detected "red sequence" and that predicted by passive evolution galaxy
models for ellipticals at the redshift estimated by the
matched-filter. The results presented in this paper show the
usefulness of color data, even of two-band data, to both tentatively
confirm cluster candidates and to select possible cluster members for
spectroscopic observations. Based on the present results, we estimate
that $\sim150$ EIS clusters with $z_I\lesssim0.6$ are real, making it
one of the largest samples of galaxy clusters in this redshift range
currently available in the southern hemisphere.   }

\maketitle

\section{Introduction}
\label{sec:intro}

Clusters of galaxies are prime targets in observational cosmology for
several reasons among which: they are the largest gravitationally bound
structures observed in the universe and as such are useful tracers of
the large-scale structure; their abundance as a function of redshift
can set strong constraints on the cosmological parameters; studies of
the properties of cluster galaxies can provide insight on their
star-formation and merging history and set valuable constraints to
galaxy formation and evolution models. However, such studies are only
possible using statistically representative samples of clusters
spanning a large range of redshifts. Unfortunately, such samples are
currently not available, with most of the existing cluster catalogs
being either limited in number or redshift coverage. In recent years,
several attempts have been made to conduct systematic searches for
more distant systems both in optical/infrared and X-ray utilizing a
large variety of detection techniques
\citep[e.g.][]{gunn86,lidman96,lumsden92,postman96,zaritsky97,gonzalez01,gladders00}.
Other promising methods include the use of mass reconstruction maps
built from weak lensing signals and the Sunyaev-Zeld'ovich effect to
conduct blind cluster searches over wide areas.

While the number of candidates has significantly increased, especially
from optical/infrared searches, the number of confirmed and
well-studied cases with $z\sim1$ remains relatively small
\citep[e.g.][]{postman01}.  Therefore, the present samples fall far
short of what is desirable for: exploring the relationship between
systems selected using different detection algorithms, a more complete
characterization of cluster properties and a better understanding of
the evolution of clusters and their galaxy population. The major
obstacles for enlarging the sample of well-studied cluster galaxies
have been a combination of several factors which include the
reliability of cluster candidates, interloper contamination and
instrumental limitations.  The reliability issue has in general
favored follow-ups of X-ray selected samples, which are currently
small and unlikely to grow significantly in the near future.
Therefore, to immediately benefit from the available large aperture
telescopes and multi-object spectrographs for galaxy cluster research,
for the time being one must rely on optical/infrared surveys such as
those being carried out by the ESO Imaging Survey (EIS) project. Using
the original $I$-band survey data
\citep{nonino99,prandoni99,benoist99} a catalog of cluster candidates
was compiled \citep{olsen99a,olsen99b,scodeggio99} comprising
$\gtrsim300$ candidates in the redshift range $0.2\leq z\leq1.3$
within an area of $\sim15$ square degrees, until recently the largest
publicly available sample of its kind in the southern
hemisphere. However, candidate clusters based on a single passband are
normally viewed with skepticism because of the known problems caused
by projection effects. In order to address this concern, in the
present paper we use the publicly available $V$-band data to
demonstrate that by using color information one can provide
independent supporting evidence of the reality of a large fraction of
the candidate clusters.

We use the available $V$ data in two complementary ways. First, we
extend the analysis carried out by \citet{olsen99b} and treat the data
from the $V$-survey as an independent data set to which we apply the
matched-filter algorithm and test the reproducibility of the
detections.  The $V$-detections are then cross-compared to those
obtained from the analysis of the $I$-band data. Second, we use the
$(V-I)$ color to search for a simultaneous spatial and color
concentration at the location of the $I$ detections. Such
concentrations are interpreted as being associated with the red
sequence observed in rich clusters over a broad range of redshifts
\citep[e.g. ][]{aragon-salamanca93,stanford98,lubin96b}. This
red sequence is interpreted as being an indication that rich clusters
contain a core population of passively evolving elliptical galaxies
which are old and coeval. This method is similar to that described by
\citet{gladders00} in the so-called red sequence cluster survey.

The $V$-band data restricts the analysis primarily to cluster
candidates with $z\lsim0.6$ since at larger redshifts ellipticals
dropout from the sample as the 4000\AA\ break moves past the
$V$-band. However, the analysis presented here illustrates the
methodology that will be employed in forthcoming papers to investigate
high-redshift candidates using the multi-color optical/infrared ($R$
and $JK_s$) imaging data assembled for a large number of EIS
candidates with $z\geq0.5$ ($\sim110$ and $\sim40$). These data will
also be combined with the results of ongoing spectroscopic
observations being conducted at the 3.6m telescope at La Silla
\citep[e.g.][]{ramella00} and at the VLT of color-selected systems in
different redshift intervals. Preliminary results based on an
admittedly small sample suggest that a significant fraction of these
color-selected systems are confirmed spectroscopically.

This paper is organized as follows. Sect.~\ref{sec:vband} briefly
describes the data used in the present analysis. Sect.~\ref{sec:mf}
presents the results obtained  from the comparison of matched-filter
detections in $V$ and $I$, from which a first estimate of the
confirmation rate is derived. Sect.~\ref{sec:colorslice} describes the
methodology employed to detect early-type cluster members and to measure
the typical color associated to these objects. This information is used
to obtain independent estimates of the expected confirmation rate and
the redshifts assigned to them.  A discussion of the results is
presented in Sect.~\ref{sec:discussion}. A brief summary of our
results is given in Sect.~\ref{sec:conclusions}.

\section{The Data}
\label{sec:vband}

In the present analysis we use the catalog of EIS candidate clusters
compiled by \citet{olsen99a,olsen99b} and \citet{scodeggio99} comprising
$\gtrsim300$ candidates over four patches of the sky covered by the
EIS-WIDE $I$-band survey conducted at the NTT. The candidates were
identified from the galaxy distribution extracted from these images and
presented in \citet{nonino99}, \citet{prandoni99} and \citet{benoist99}. 

In this paper the $I$-data are complemented with those taken in
$V$. The $V$-data consists of two sets of images. One taken at the NTT
with EMMI, as part of the original EIS-WIDE survey which covered
two~square~degrees (Patches A and B). The other consists of images
taken at the MPG/ESO 2.2~m telescope with ESO's Wide-Field Imager
(WFI) as part of the so-called Pilot Survey. The latter set covers
a total area of 12~square~degrees in $V$-band, out of which
6~square~degrees are publicly available.  Galaxy catalogs have been
extracted from these $V$-images and associated with those in $I$ to
produce $(V-I)$ color catalogs utilized below, covering a total area
of 8~square~degrees, or about half the area of the $I$-survey. Within
this area there are 168 $I$-band cluster candidates, corresponding to
55\% of the entire EIS candidate sample.

\begin{figure} \center
\resizebox{0.9\columnwidth}{!}{\includegraphics{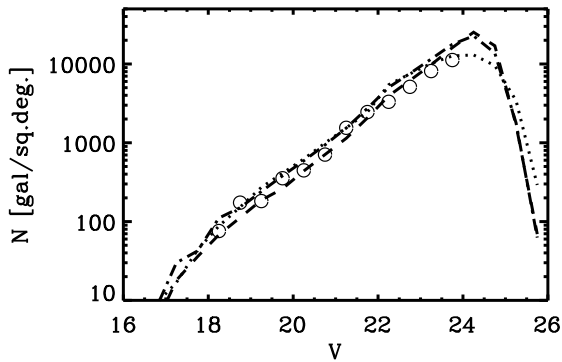}}
\caption{Galaxy number counts derived from the $V$-band images obtained
for EIS Patches~A (dashed-dotted line) and B (dashed line) and the Pilot
Survey (Patch~D, dotted line). For comparison the counts of
\protect\citet{arnouts95} (open circles) are also shown.}
\label{fig:ncounts_v} \end{figure}

Fig.~\ref{fig:ncounts_v} compares the galaxy counts obtained from the
$V$-band catalogs extracted from the images taken using different 
telescope/detector setups. Even though reaching different limiting
magnitudes, the counts are consistent with each other and in good
agreement with those of other authors \citep{arnouts95}. This
demonstrates the uniformity of the galaxy catalogs, which are not
distinguished in the analysis below. The $5\sigma$-limiting magnitudes
are estimated to be $V\sim24$ and $V\sim24.6$ \citep{dacosta00a} for the
EIS-WIDE (patches~A and B) and Pilot Survey data (patch~D),
respectively. Therefore, in the analysis presented below  we only consider
galaxies brighter than  $V=24$. 

\section{Searching for clusters in $V$-band}
\label{sec:mf}

\subsection{Matched-filter Detections}
\label{sec:mf_det}

The EIS cluster search was using the matched-filter algorithm
originally proposed by \citet{postman96} and used to build the Palomar
Distant Cluster Survey (PDCS).  This method uses a filter built
from the luminosity function and radial profile of nearby clusters. At
each position of the survey area the probability of the presence of
a cluster matching the filter is computed. By fine-tuning the filter
to match the apparent magnitudes and extent of the cluster at a series
of redshifts ($0.2\leq z\leq1.3$) a likelihood curve is
constructed. The cluster candidate is then assigned the redshift for
which the maximum likelihood is found.

In contrast to the PDCS the EIS cluster catalog was built exclusively
from matched-filter detections in the $I$-band for which two sets of
images, contiguously covering the regions of interest, were
available. These sets were treated separately to construct two galaxy
catalogs which were called ``odd'' and ``even'' catalogs, to indicate
the image mosaic from which they were extracted. The matched-filter
algorithm was run on each catalog and the resulting cluster detections
were compared. The detections were considered to be: ``robust'' if
they were detected at $4\sigma$ in one or both catalogs, or if they
were detected at the $3\sigma$-level in both the ``even'' and ``odd''
catalogs, and ``poor'' if they were detected at the $3\sigma$-level in
only one of the catalogs. These definitions were introduced by
\citet{olsen99a} \citep[see also ][]{scodeggio99} and are used below.

The $V$ images available represent another sampling of the galaxy
distribution in the region and as such can be used to define another set
of cluster candidates. In applying the matched-filter procedure to the
galaxy catalogs extracted from the $V$-band images, we use the same
cluster model as in $I$. This model assumes a Schechter luminosity
function and a radial distribution of galaxies given by a truncated
Hubble profile \citep[for more details see ][]{olsen99a}. Following
\citet{postman96} we adopt the Schechter parameters $M^*_V=-21.02$ and a
faint end slope of $\alpha=-1.1$, valid for local clusters. For the
radial profile we adopt a core radius of $r_\mathrm{c}=100h^{-1}\mathrm{kpc}$ and
a cut-off radius of $r_\mathrm{co}=1h^{-1}\mathrm{Mpc}$. We use a cosmological
model with $H_0=75\mathrm{km/s/Mpc}$, $\Omega_0=1.0$.  

In the discussion presented below it is important to have a rough idea
of the redshift limit for cluster detection in each of the passbands.
This can be obtained from the limiting magnitudes of the input galaxy
catalogs ($I_\mathrm{lim}\sim23$ and $V_\mathrm{lim}\sim24$, Vega system), by
assuming that the cluster population is formed predominantly by
early-type galaxies, and by assuming a particular model for the
evolution of the stellar population. The maximum contrast of a cluster
over the field population occurs in the range of magnitudes
$m^*\pm1\mathrm{mag}$, where $m^*$ is the apparent Schechter magnitude
of the cluster.  Therefore, the limiting redshift corresponds
approximately to that for which $m^*+1$ is comparable to the limiting
magnitude of the galaxy catalog. Assuming a no-evolution model
this would, for the data considered here, imply a redshift limit of
$z\sim0.6$ and $z\sim1.1$ for the $V$- and $I$-candidates,
respectively. If one instead assumes a passive-evolution model the
redshift limits would become slightly larger. As it will be seen
below, $z\sim0.6$ as determined for the $V$-band is indeed the
practical limit of the present analysis.

\begin{figure*} \center
\resizebox{\columnwidth}{!}{\includegraphics{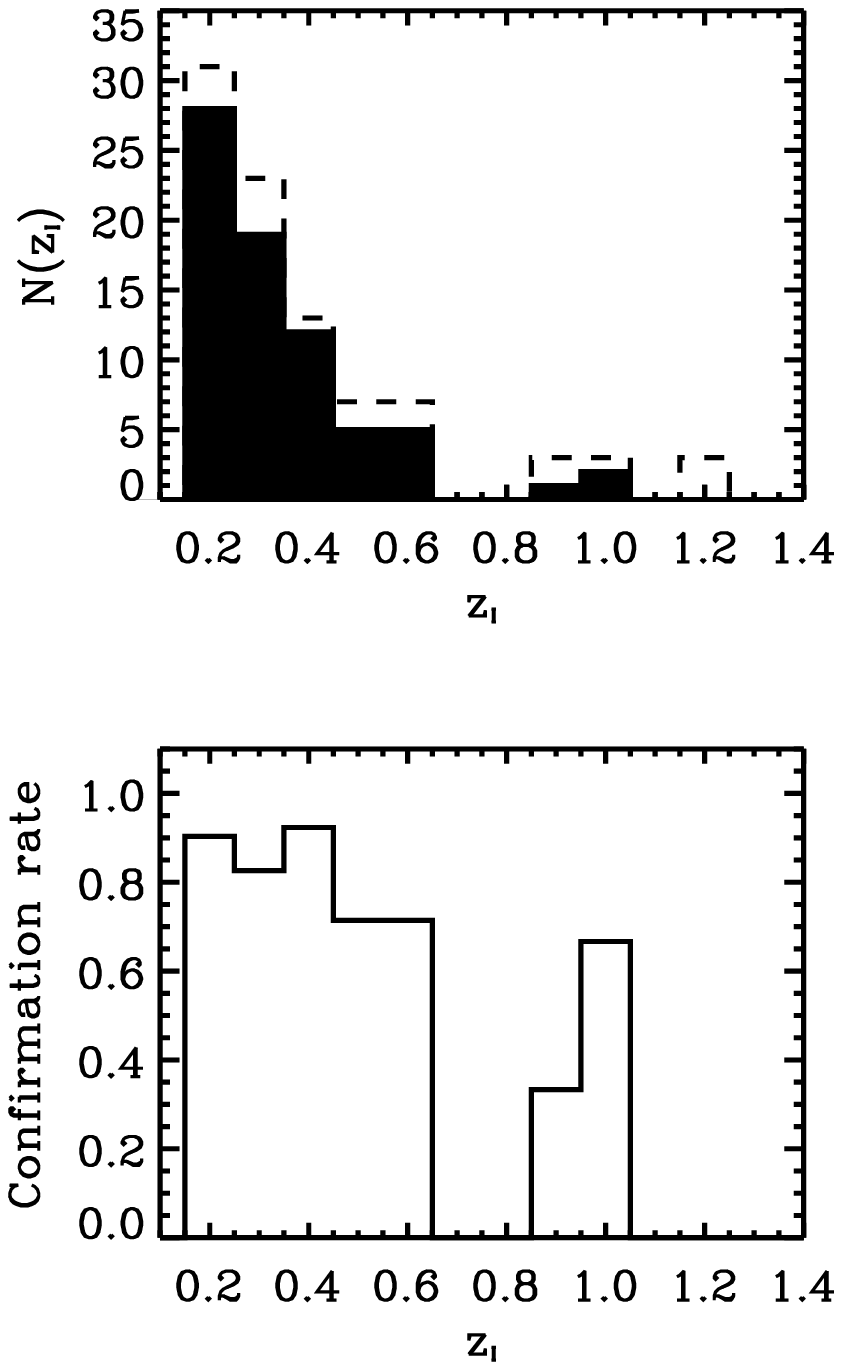}}
\resizebox{\columnwidth}{!}{\includegraphics{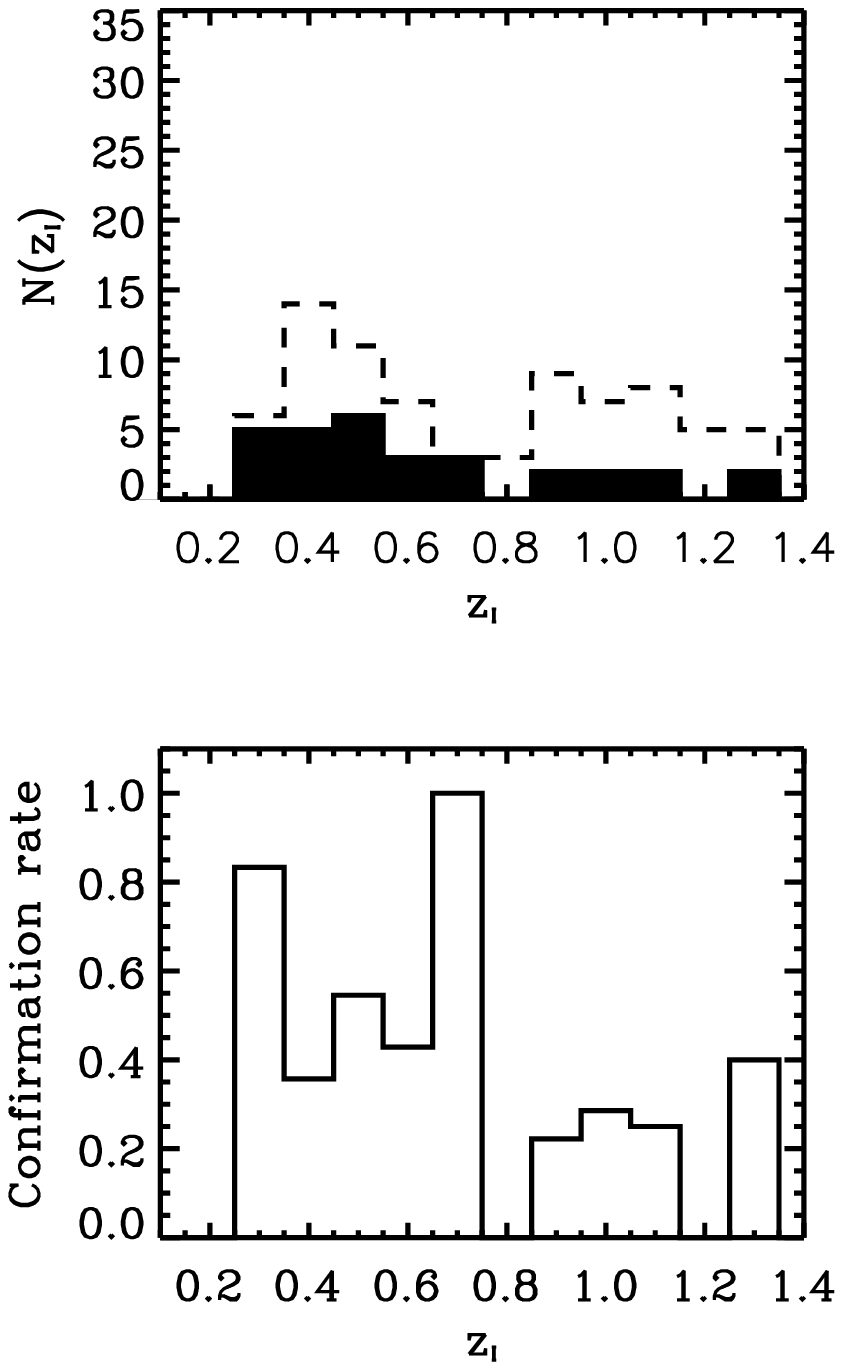}}
\caption{The matching rate as function of redshift for the ``robust''
(left column) and ``poor'' (right column) candidates as explained in the
text. The upper panel shows the redshift distribution for all the
candidates (dashed  histogram) and those with counterparts in the
$V$-band (filled histogram). The lower panel shows the respective
matching frequency as function of redshift which is interpreted as  an
estimate of the confirmation rate of the original $I$-detections.}
\label{fig:match_rate} \end{figure*}

\subsection{Comparison between $V$- and $I$-detections}
\label{sec:compmf}

The cluster candidates identified by applying the matched-filter
algorithm to the galaxy catalogs extracted from the $V$ images are
compared to the $I$-band candidates. Below we only consider those
which match $I$-candidates, even though there are also $V$-candidates
without corresponding $I$-detections ($\sim40$, which will be
discussed elsewhere). Detections in the different passbands are
matched by considering only positional coincidence. For this we
require the angular separation to correspond to a projected separation
of less than $0.5h^{-1}\mathrm{Mpc}$ at the redshift estimated for the
$I$-candidate ($z_I$), regardless of the matched-filter redshift
estimate, $z_V$, based on the $V$ catalogs. In the following we
investigate the frequency of matches and compare the $V$- and
$I$-estimated redshifts.

Fig.~\ref{fig:match_rate} summarizes the results of the
cross-comparison between $V$- and $I$-candidates. The top panels show the
redshift distribution of the matched candidates compared with the
redshift distribution of the sub-sample of EIS $I$-candidates located
within the area covered by $V$. The left panels refer to ``robust''
candidates (90 candidates) and the right to the ``poor'' ones (78
candidates).  The overall matching rate is 61\% (102 out of 168 $I$-band
candidates), and varies with redshift as shown in the lower panels of
Fig.~\ref{fig:match_rate}.

From Fig.~\ref{fig:match_rate} one sees that for $z_I\leq0.6$ most
clusters are detected in both passbands.  In this redshift interval
the matching rate (hereafter interpreted as an estimate of the
confirmation rate) for the whole sample is $\sim75\%$ (88 out of
118). Additionally restricting the sample to ``robust'' candidates, the
confirmation rate is higher, reaching $\sim82\%$ (69 out of 81).  It
is also interesting to note that about 50\% of the ``poor'' candidates
in this redshift range are associated with $V$-detections, showing an
above average rate at $z_I=0.3$.  This could be due to low-redshift
poor clusters detected at a low significance level in a
single-passband.

At larger redshifts ($z_I\geq0.7$) only 14 ($\sim26\%$) of the 54
$I$-band cluster candidates are matched by a $V$-detection. This lower
confirmation rate is not surprising considering that for systems with
$z\gsim0.6$ the 4000\AA\ break lies redwards of the $V$-band.  Note,
however, that we find matching $V$-detections over a broad range of
redshifts including some very high-redshift systems ($z_I\gsim1$),
especially in the "poor" sample. In order to investigate the nature of
these matches the 14 cases of clusters with $z_I>0.6$ and with matched
$V$ detections were visually inspected. In five cases ($z_I=0.7, 0.7,
0.9, 1.1, 1.3$) the $V$-detection is clearly associated with a
foreground concentration unrelated to the $I$-band cluster candidate,
even though the $I$-band image does appear to include a cluster
candidate. In one case ($z_I=1.3$) the $V$-image contains a satellite
track which leads to spurious galaxy detections forming an artificial
concentration of objects in the $V$ catalog. For the remaining eight
cases the $V$-band images do show concentrations of galaxies
consistent with those seen in $I$. These eight candidates - four
robust and four poor $I$-detections - have redshifts in the range
0.7-1.1. As mentioned above, while early-type galaxies are not
expected to be detected in this redshift range, the cluster detection
based on the matched filter technique takes into account the entire
galaxy population. Therefore, this makes this method less sensitive
than others (see below) to the fading in the $V$-band of the
early-type galaxies.  One should also keep in mind the large
uncertainties in the redshift estimates at high redshift, which could
lead to the matching $VI$-detections with large differences in
redshift estimates. However, as shown below this does not seem to be
the case. Of course, there is still the possibility of the
superpositions of clusters at different redshifts with $I$ detecting a
distant richer system, while $V$ detecting a foreground one. Deciding
among these possibilities will have to await the results of
spectroscopic observations. Note that a comparable fraction of $VI$
detections in the redshift interval being considered was found by
\citet{postman96} \citep[see also ][]{lubin96b}.

\begin{figure}
\center
\resizebox{0.9\columnwidth}{!}{\includegraphics{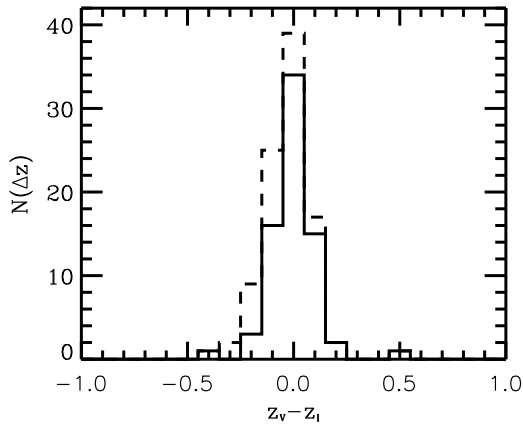}}
\caption{The distribution of differences of redshifts estimated from
the $V$ and $I$ matched-filter identifications for the entire sample
(dashed line) and the ``robust'' candidates (solid line).}
\label{fig:dz_dist}
\end{figure}

\subsection{Redshift Comparison}

Besides confirming $I$-candidates, the $V$-detections can also be used
to assess the reliability of the original cluster redshift estimates
based on the application of the matched-filter algorithm to the
$I$-band galaxy catalog. Fig.~\ref{fig:dz_dist} shows the distribution
of redshift differences estimated by the matched-filter algorithm for
associated $V$- and $I$-detections. The figure shows this distribution
for the whole sample, considering matches over the entire redshift
range, but eliminating the six spurious matches discussed in the
previous section.  Also shown is the distribution of redshift
differences for the sample of ``robust'' candidates.  Taking the
sample as a whole, we find a mean difference of $\langle
z_V-z_I\rangle=-0.03$ and a scatter of $0.12$. For the ``robust''
candidates the mean difference is significantly smaller $\langle
z_V-z_I\rangle=-0.003$ with a scatter of $0.11$.  Interestingly, all
the eight cases with matched $VI$ detections and redshifts $z_I>$0.6,
discussed in the previous section, have consistent redshifts within
the estimated errors, giving further credence to their
confirmation. Note that two of the ``robust'' cases are outliers - one
with $(z_V-z_I)=0.5$, $z_V=0.7$ and $z_I=0.2$; another with
$(z_V-z_I)=-0.4$, $z_V=0.2$ and $z_I=0.6$. To investigate these cases
we visually inspected the images and the ``even'' and ``odd''
detection lists (see Sect.~\ref{sec:mf_det}).  From this inspection we
conclude that: in the first case the cluster seems real in both
bands. However, the original $I$-detection had two different redshifts
in the ``even'' ($z_I=0.4$) and ``odd'' ($z_I=0.2$) catalogs. The
latter was included in the EIS cluster candidate catalog, because it
had a higher significance. Clearly, for this cluster $z_I$ is
uncertain, but there is little doubt about the reality of the
cluster. In the second case the redshift discrepancy is probably due
to the superposition of a low redshift system detected in $V$ and a
more distant cluster detected in $I$. Finally, we emphasize that the
above results also demonstrate the reliability of the redshift
estimates and can be used for selecting sub-samples in different
redshift intervals, if an error of about 0.15 is properly taken into
account.

\section{Searching for cluster early-types}
\label{sec:colorslice}

Confirmation of candidate clusters by matching $V$ and $I$ relies only
on the fact that the $V$ images and the derived galaxy catalogs may
suffer different effects than those in $I$-band. Thus, the
reproducibility of the results  obtained from data in different
passbands serves to corroborate the original findings, thereby
increasing the significance of the detections. This test, however,
relies on the same underlying assumptions of the matched-filter method -
clusters are nearly spherical, have the same projected density profile
and are characterized by a Schechter luminosity function, similar to the
one observed in local clusters.

An alternative way of searching or checking the reality of clusters is
to use the additional information provided by the $(V-I)$ color. One
way is to search, at the location of the $I$ detections, for
early-type galaxies which are known to form a tight color-magnitude
relation and to populate the central regions of galaxy
clusters. Another is to conduct blind searches for clustering of
early-type galaxies with colors consistent with their being at
different redshift intervals. These methods make no assumption about
the global properties of clusters but have to rely on galaxy evolution
models. These color-based methods are complementary to the
matched-filter approach.  In addition, splitting galaxy catalogs in
color intervals leads to an increase in the contrast between the
cluster and the background. This enhanced contrast can be particularly
important for the confirmation of poor cluster candidates.
Consistency between the two approaches, for reasonable models of
galaxy evolution, can thus lend further support to the reality of a
cluster candidate. Additionally, an important advantage of color-based
methods is that they provide a sample of candidate early-type cluster
members which can be used for follow-up spectroscopic observations.

\subsection{Color-slicing procedure}

\begin{figure*}
\center
\resizebox{0.9\textwidth}{!}{\includegraphics{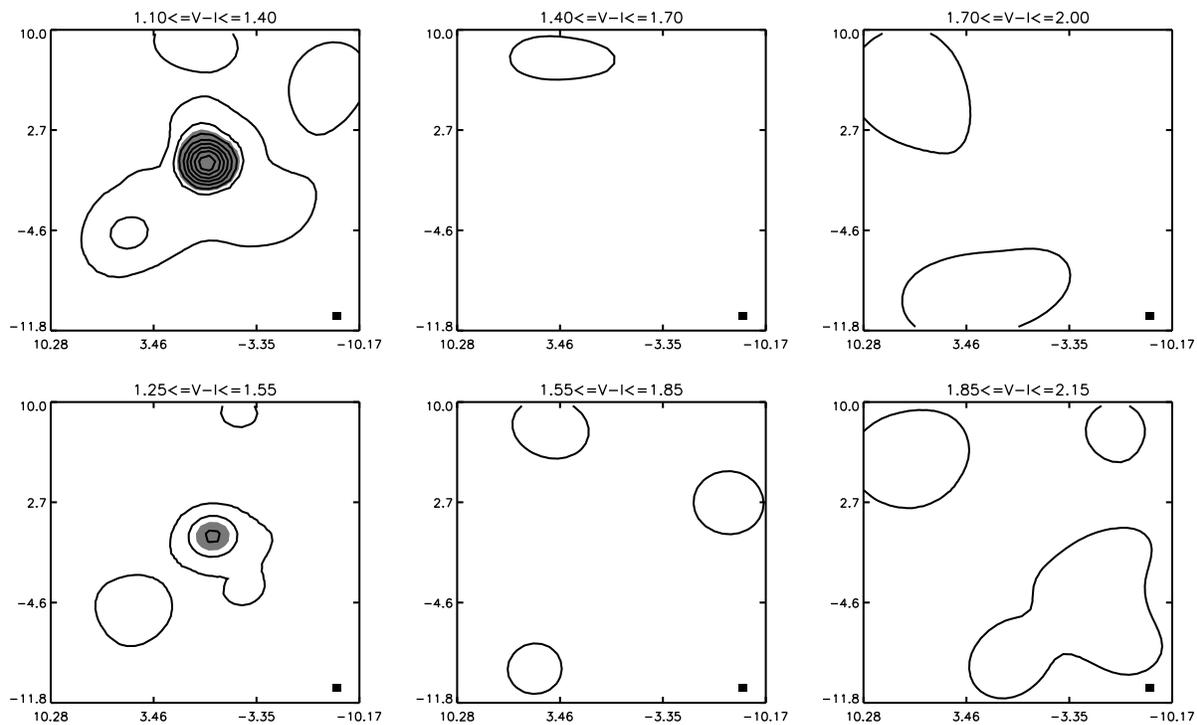}}
\caption{Six density maps for one cluster candidate. The upper and
lower row each correspond to the first three maps of each of the
subsets of seven density maps. The remaining eight slices do not contain
relevant information in this particular case. The two sets are shifted
by half the width of a color slice to make sure that a cluster is not
split between two slices, whereby it would be less likely to be
detected. The contours mark the regions with densities in excess of
the mean density and are separated by one~$\sigma$. The shaded areas
mark the significant pixels. The scale on the axes is arcmin. In each
panel the adopted pixelscale is marked as the filled square in the
lower right corner. }
\label{fig:dens_maps1}
\end{figure*}

In order to detect clustering of early-type galaxies associated to the
$I$-detections we search for simultaneous overdensities in color and
projected density distribution.  We select galaxies within a region of
$3\times3h^{-2}\mathrm{Mpc^2}$ and within a magnitude interval of
$m^*\pm2$, where $m^*$ is the expected apparent Schechter magnitude,
around the $I$-candidate position and split the galaxy sample in color
intervals $\Delta(V-I)$=0.3, covering the range
$1.1\leq(V-I)\leq3.35$ corresponding to the redshift range $0\lsim
z\lsim1.2$. For each color interval we construct a smooth density map
using an adaptive kernel smoothing \citep[see ][]{merritt94}. The
pixel scale of the density map scales with the estimated redshift of
the candidate considered. In the present analysis it is chosen to be
$75h^{-1}\mathrm{kpc}$, which roughly corresponds to the cluster model
core radius. We select significant pixels at the $3\sigma$ confidence
level relative to the field galaxy density distribution in the color
interval being considered. The density peaks are then identified in
the spatial distribution of these pixels. The background is computed
averaging the smoothed density distributions of 60 randomly selected
areas chosen within the EIS patches.  If a significant peak lies
within a radius of $375h^{-1}\mathrm{kpc}$ (corresponding to 5~pixels)
from the nominal cluster position we consider it to be a detection of
the cluster candidate red sequence. We interpret these detections as a
confirmation of the $I$-detection. The search radius adopted allows
for possible offsets of early-type galaxies relative to the cluster
center as defined by the matched-filter position. In cases where
significant pixels are found in more than one color interval the color
assigned to the cluster red sequence is that of the slice in which the
largest detected overdensity is found. In the case of superpositions
in the cluster fields we manually assign the color that is closest to
that predicted by galaxy evolution models.

Fig.~\ref{fig:dens_maps1} illustrates the results of the color-slicing
analysis showing two sets of density maps obtained in different color
intervals. In this particular case, significant pixels are detected in
two different slices. The highest density is obtained in the slice of
$1.10\leq(V-I)\leq1.40$, thus a color $(V-I)$=1.25 is assigned to the
cluster.  More details about the color slicing procedure adopted here
can be found in \citet{olsen00}. Here it suffices to say that most of
the original $I$-candidates with $z_I\lsim0.4$ exhibit well-defined
red-sequences which are clearly visible in the $(V-I)\times I$
color-magnitude diagram. Furthermore, the characteristic color of the
observed red sequence is, in general, consistent with that derived by
the automatic procedure described above. For clusters with $z_I\gsim0.4$
the data are not sufficiently deep for the red-sequence to be clearly
visible for individual clusters (\eg Lubin 1996) and we must rely on the
color-slicing detections.

\begin{figure*}
\begin{center}
\resizebox{\textwidth}{!}{\includegraphics{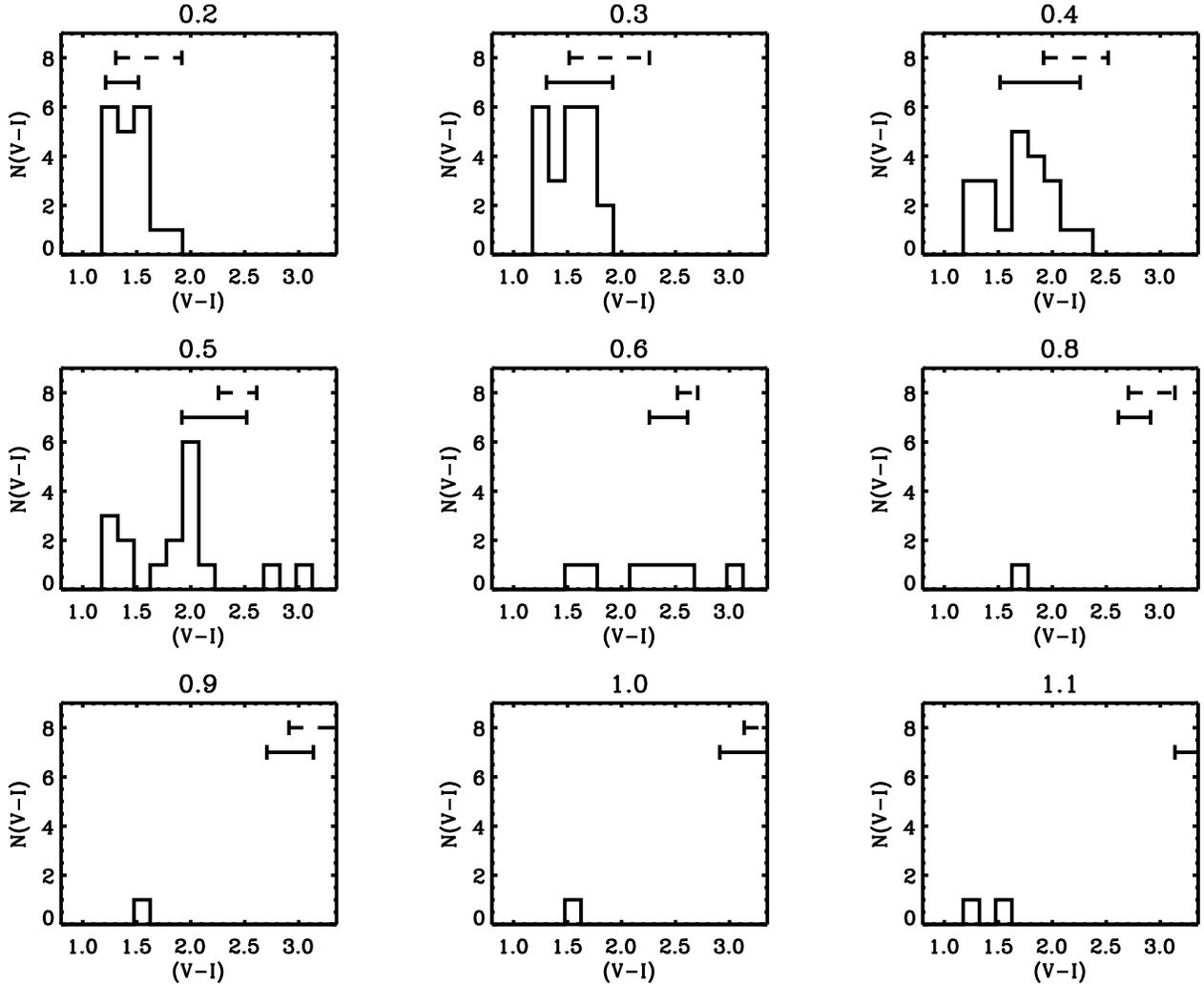}}
\end{center}
\caption{Each panel shows the histogram of $(V-I)$-colors assigned to
the cluster candidates with estimated redshift as indicated above each
panel. The dashed bars indicate the $(V-I)$-interval corresponding to
a redshift range of $z\pm0.1$ corresponding to the estimated
uncertainty. The solid bars mark the $(V-I)$-interval centered at
$z-0.1$ as discussed in the text. The colors are estimated by a
passive evolution model for the stellar population as described in the
text.}
\label{fig:color_hist}
\end{figure*}

\begin{figure*}
\begin{center}
\resizebox{\columnwidth}{!}{\includegraphics{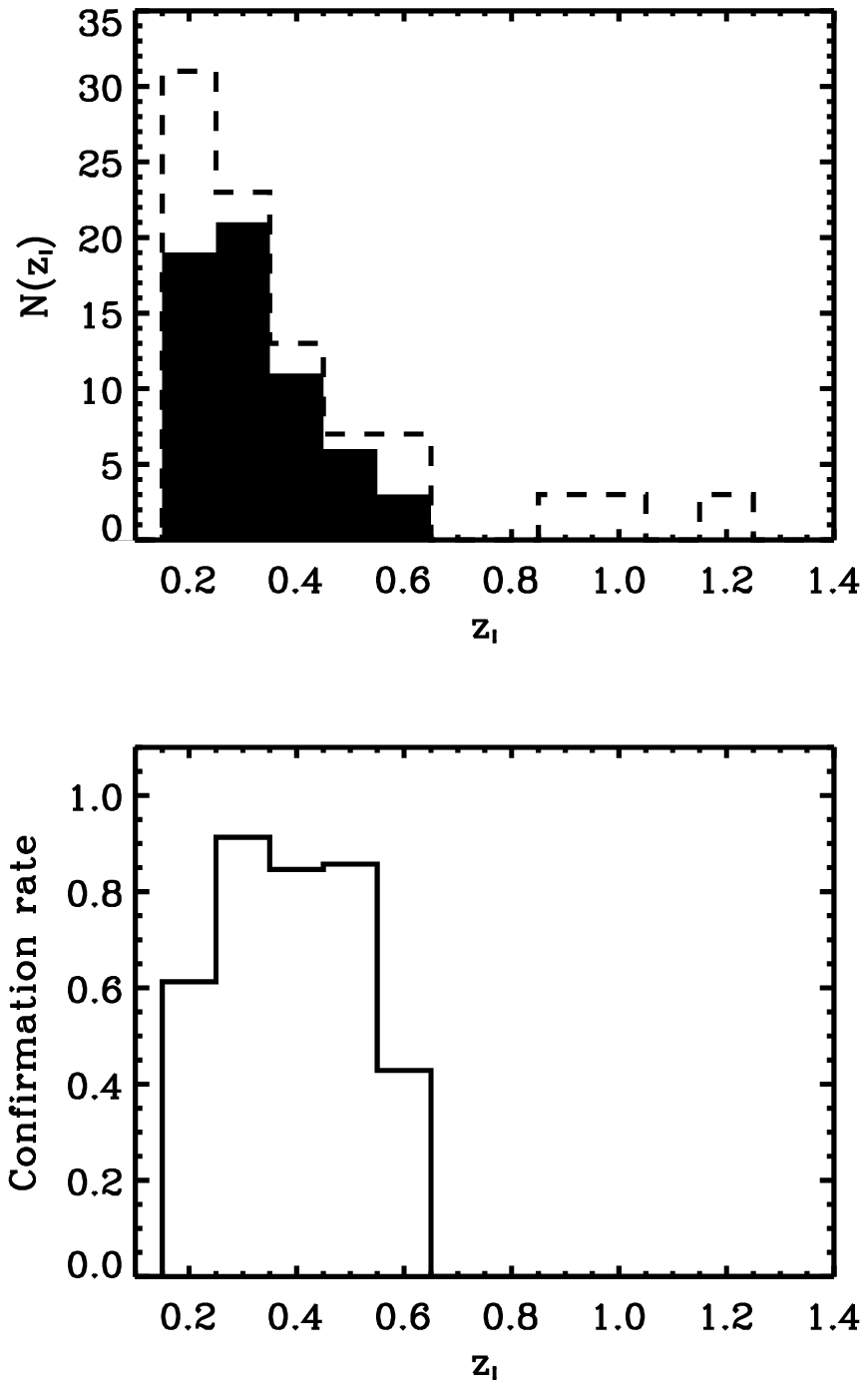}}
\resizebox{\columnwidth}{!}{\includegraphics{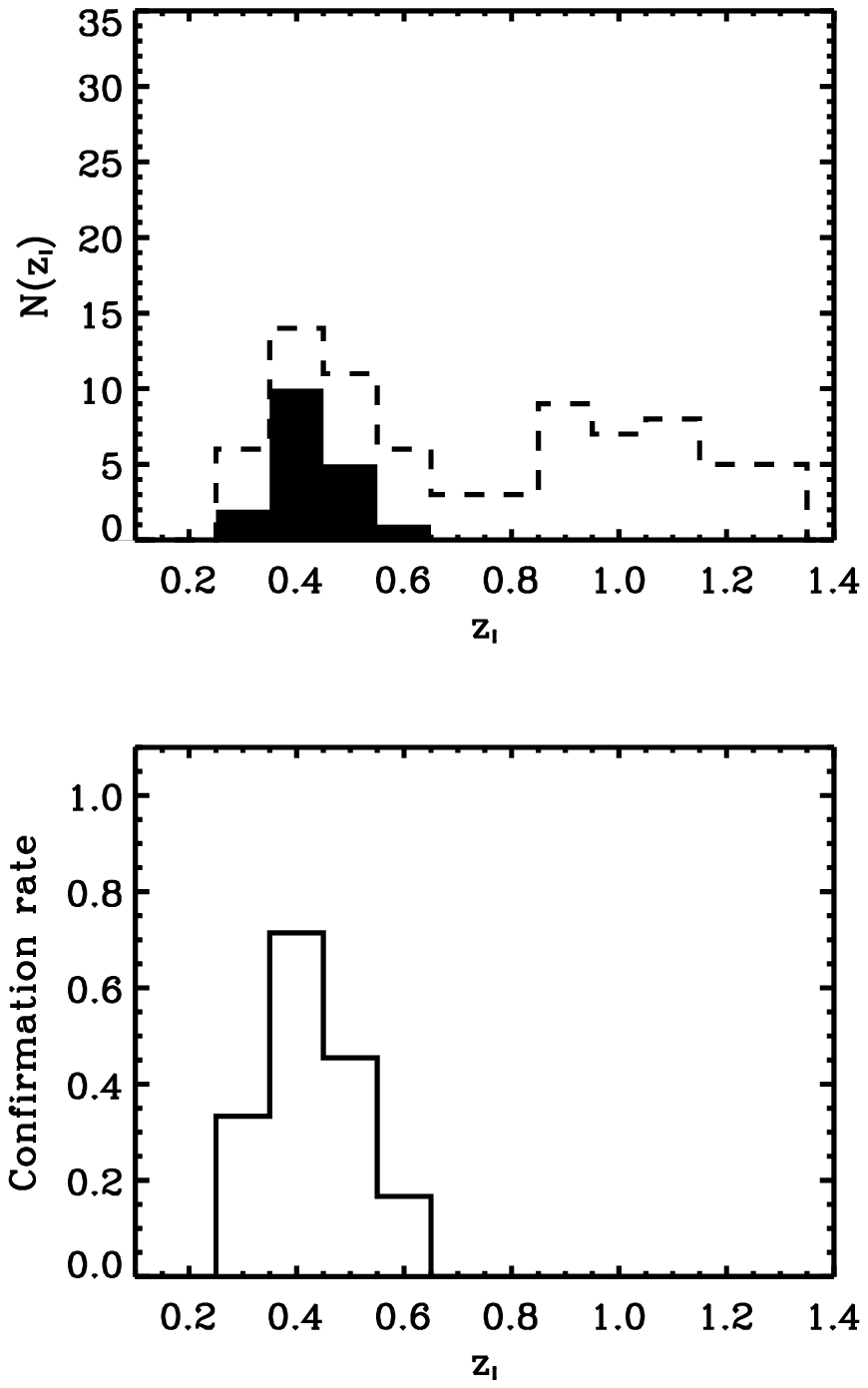}}
\end{center}
\caption{The panels to the left show the results for the sample of
``robust'' candidates and the right column for the ``poor'' ones. Upper
panels: The redshift distribution of clusters detected by the color
slicing method (dark histogram) compared to the distribution for all
candidates with $V$- and $I$-data (dashed line). Lower panels: The
confirmation rate computed as the number of candidates with a color
slice detection (the dark area in the upper panel) relative to the
total number of candidates with $V$- and $I$-data (the dashed line in
the upper panel.)}
\label{fig:vi_z_conf_hist}
\end{figure*}

\subsection {Detection of the red sequence}

Applying the color-slicing technique described above to the $I$-band
candidates we find significant detections for 94 out of 168
($\sim56\%$) candidates, a value comparable to that previously found
for matched $V$- and $I$-detections. The redshifts of the candidates
cover the range from $z_I=0.2$ to $z_I=1.1$. Of the candidates with
significant detections 69 were also detected by the matched filter in
$V$, which correspond to 68\% of the 102 candidates detected in both
$V$ and $I$ by the matched filter. When considering only the cases
with $z_I\leq0.6$ for which the $V$-band filter is most suitable, we
find that 79\% of the $V$ and $I$ matched-filter detections are
confirmed by the color slicing. In order to investigate the colors
assigned to each candidate we show in Fig.~\ref{fig:color_hist} the
histograms of the colors assigned to candidates in each redshift bin.
Redshift bins without color-slice detections are not shown. The
intervals shown on the top of each panel indicate the uncertainty in
the $(V-I)$-color due to errors in the redshift estimates.  In this
plot the $(V-I)$-color interval is computed using a model for a
passively evolving stellar population for a redshift range of
$z\pm0.1$ (dashed line).  The adopted model assumes a Salpeter initial
mass function and a burst of star formation lasting less than 1~Gyr
and an epoch of formation corresponding to $z_\mathrm{f}=10$ (see further
discussion below). One immediately notices that the predicted colors
are redder than those obtained for the cluster candidates. This is
consistent with results based on simulated data showing that the
matched-filter procedure overestimates the redshift
\citep{olsen00}. In Fig.~\ref{fig:color_hist} we also show the color
interval (solid line) predicted by the model taking this zero-point
offset into account (see below). Note that the $(V-I)$-uncertainty
becomes smaller at $z\sim0.5-0.6$, because at these redshifts the
evolution of the color with redshift is slow.

From the figure one finds that in the redshift range $z_I\sim0.2-0.6$
the color assigned to the cluster red-sequence form a well-defined
distribution consistent with model predictions if, as mentioned above,
one assumes that the redshift estimates, based on the matched-filter,
are indeed overestimated.  However, some outliers are clearly present,
their number increasing with redshift. 

In the redshift interval $z_I\sim0.5-0.6$, we find seven blue and
three red outliers. In order to investigate their nature the images
and density maps for these detections were individually inspected. In
all cases the position of the detections correspond well with that of
the cluster candidate. Moreover, for all of the blue outliers but one,
the color assigned by the automatic procedure is associated to
foreground detections.  However, in most of these cases redder
peaks are present but at lower significance, preventing them from
being considered detections by the automatic procedure. These redder
peaks are probably related to the cluster red sequence, and thus
represent the real color of the early-type members.  In this paper,
we take a conservative stance and do not consider these cases as true
detections and discard them from any further analysis.  Therefore, we
conclude that the increase in the number of blue outliers is due
primarily to the fading of early-type galaxies as the 4000{\AA} break
moves through the $V$-band filter and that the size of the search
radius adopted has no effect on the number of outliers.  This effect
is unavoidable and its onset depends both on the redshift of the
cluster and on its nature (\eg composition, number of bright
early-type galaxies). There are two possible ways of further
characterizing these clusters. One is to consider the spatial
distribution of $V$-dropouts, the other to use $R$-band data collected
for this project. This will be considered in a separate paper.  It is
worth noting that in the redshift interval $z_I\sim0.5-0.6$, we also
find three red outliers, all of which are consistent with the
interpretation that the red sequence of the cluster has indeed been
detected.

At $z_I\geq0.7$ we find five candidates exhibiting a color slice
detection. Since at these redshifts the 4000{\AA} break has completely
moved through the $V$-band filter we expect that most of the
early-type galaxies have dropped out from the sample and that no
red sequence can be detected. In fact, from the inspection of the
images and density maps we conclude that all the detections are due to
foreground concentrations. This conclusion is also valid for the two
cases with matching $V$ and $I$ matched-filter detections ($z_V$=0.7,
$z_I=0.8$; $z_V$=0.9, $z_I=1.0$). While these systems are detected by
the matched-filter technique, which relies on the overdensity of the
entire cluster galaxy population, they would not be detected based on
the the color-slicing method for reasons mentioned above.
      
In Fig.~\ref{fig:vi_z_conf_hist} we compare the
redshift distributions of the $I$-detections with those for which a
red sequence has been detected. As before, we show the sub-samples
comprising ``robust'' and ``poor'' candidates, separately. The lower
panels show the estimated confirmation rate as a function of the
matched-filter redshift $z_I$ derived for the $I$-candidates.

For $z_I\leq0.6$ we find 80 clusters confirmed out of 118
($\sim68\%$).  Restricting the sample to only ``robust'' candidates we
confirm 60 out of 81 candidates ($\sim74\%$). In this redshift range
we find that the confirmation rate is nearly constant, except for the
outermost bins, both showing lower confirmation rates. We point out
that for very nearby systems $z_I\sim0.2$, the red-sequence may not be
detected automatically due to the large angular size of the
clusters. For the $z_I\sim0.6$ the lower rate possibly reflects the
fading of early-type galaxies due to the redward shift of the
4000{\AA} break. Finally, note that for $z_I\lsim0.6$, $\sim50\%$ of
the ``poor'' candidates are confirmed by the color-slicing method, in
agreement with our earlier findings based on matched $V$- and
$I$-detections.

Clearly, the color-slicing method, which is based on the identification
of elliptical galaxies, is more sensitive to their fading in the
$V$-band than the method that relies on the detection of matched $V$-
and $I$-detections which depend on the galaxy population as a
whole. This restricts the effectiveness of the identification of red
sequences up to roughly $z\sim0.6$ and thus explains the sharp drop
observed in the confirmation rate beyond $z_I$=0.6. This effect
combined with variations in individual cluster properties, the
uncertainties in the redshift estimate and colors, and projection
effects might also explain the significant scatter in the estimated
color (discussed in the next section) of the red sequence for systems
with $z_I\gsim0.5$.

\begin{figure}
\begin{center}
\resizebox{0.9\columnwidth}{!}{\includegraphics{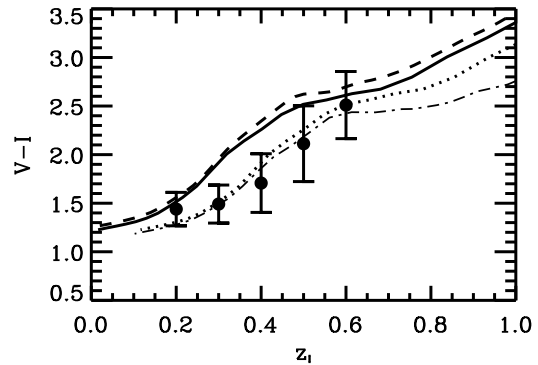}}
\end{center}
\caption{The relation between the assigned $(V-I)$-color and estimated
redshifts for the clusters that are considered confirmed.  The filled
circles mark the mean color in each redshift bin with $1\sigma$-error
bars. The dashed line is the color evolution for a non-evolving
elliptical galaxy (only applying the appropriate K-corrections for the
redshifted galaxy spectrum), and the solid line indicate the color
evolution for a passively evolving galaxy with $z_\mathrm{f}=10$
\citep{bruzual93}. The dotted line shows the locus of the passive
evolution shifted by $\Delta z=+0.1$. The dot-dashed line shows the
locus of the passive evolution of a galaxy formed at $z_\mathrm{f}=2$ shifted
by $\Delta z=+0.1$.}
\label{fig:final_vi_z}
\end{figure}

\subsection{Colors of early-type cluster members}

The color-slicing method provides an estimate of the color of the
detected red sequence of a cluster candidate, which can be used to
check the consistency with that predicted by a particular galaxy
evolution model at the estimated redshift $z_I$. This can be used
either to constrain the epoch of formation of ellipticals or inversely
to convert a measured color to an independent redshift estimate, as
discussed below.

Fig.~\ref{fig:final_vi_z} shows the mean colors and the 1~$\sigma$
error bars of the cluster red sequence, as assigned by the
color-slicing method, as a function of $z_I$. For comparison, we also
show the colors predicted for non-evolving \citep[dashed line, using
spectra from ][]{kinney96} and passively evolving \citep[solid line,
][]{bruzual93} ellipticals formed at $z_\mathrm{f}=10$ as a function of $z$.
As also seen in Fig.~\ref{fig:color_hist} one immediately finds that
the matched-filter estimated redshifts overestimate the true redshift
by about 0.1 in order to match the colors of present-day
ellipticals. Instead of correcting the $z_I$ values we shift the curve
corresponding to the passive evolution model to take into account this
zero-point offset (dotted line). We also include a model for passively
evolving galaxies, adopting a formation redshift of $z_\mathrm{f}=2$, already
corrected for the zero-point offset mentioned above (dot-dashed line).
It is worth pointing out that the model curves are neither very
sensitive to the adopted cosmological model nor do the predicted
colors of ellipticals vary significantly in the redshift range
considered here. Currently, several studies put a lower limit on the
formation epoch of ellipticals of $z_\mathrm{f}\sim2$.  From the figure it can
be seen that we find good agreement between the measured mean and
predicted colors. The slightly larger scatter at $z\gsim0.4$ is
probably due to larger errors in the matched-filter estimated errors
and to the effect of the 4000{\AA} break being shifted into the
spectral region of the $V$-band filter.

\subsection{Redshift comparison}

\begin{figure}
\begin{center}
\resizebox{0.9\columnwidth}{!}{\includegraphics{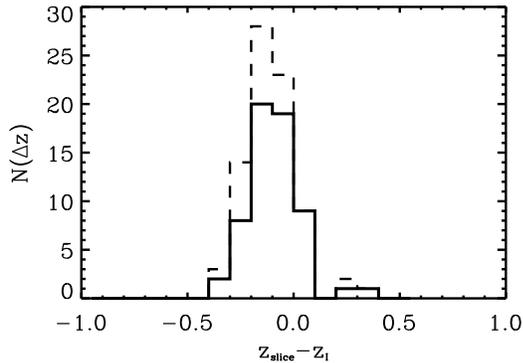}}
\end{center}
\caption{The distribution of redshift differences between color
estimated redshifts and the matched filter estimated redshifts. The
color estimated redshifts are based on the \protect\citet{bruzual93}
passive evolution scenario explained in the text. The dashed line
shows the results for the sample as a whole while the solid line show
the ``robust'' sample results. }
\label{fig:zvi}
\end{figure}

Given the overall agreement between model predicted colors and those
estimated from the color slicing procedure, the $(V-I)$-color assigned
to the red sequence of cluster early-types can, alternatively, be used
to obtain an independent redshift estimate for those clusters that are
confirmed. Using the passive evolution model introduced earlier, we
convert red-sequence colors into redshifts. A comparison between these
redshifts and those based on the matched-filter is shown in
Fig.~\ref{fig:zvi}, where the distribution of the differences between
these redshift estimates is presented. We find that in the mean the
matched filter overestimates the redshift by $\Delta z\sim+0.11$ with
a scatter of $\sigma(\Delta z)\sim0.12$ both for the ``robust'' and the
entire sample. The outliers with $(z_\mathrm{slice}-z_I)\sim0.3-0.4$
corresponds to the red outliers in Fig.~\ref{fig:color_hist}.

\section{Discussion}
\label{sec:discussion}

In this paper we have argued that direct multi-band imaging data can
be used to provide supporting evidence for the reality of candidate
clusters of galaxies derived from the application of the
matched-filter algorithm to a single band galaxy catalog. To
illustrate this point we have, in this paper, used publicly available
$VI$ data to compare $V$ and $I$ detections and to use the $(V-I)$
color to identify simultaneous concentrations in space and color,
possibly reflecting the existence of galaxies along the
color-magnitude relation observed in real clusters. The nature of the
two methods is significantly different both regarding their underlying
assumptions, the nature and methodology of detections and the
resulting estimates of the redshifts. The matched filter assumes that
clusters are spherical concentrations following a specific density and
luminosity distribution, while the color-slicing method relies on the
fact that in real clusters early-types are located preferentially in
the central regions and form a tight color-magnitude
relation. Similarly, the redshift estimates are independent, one
relying primarily on magnitudes while the other on colors.

Our main results can be summarized as follows.  For $z\leq0.6$ the
overall confirmation rate by either method is high $\sim65\%$ for the
color-slicing and $\sim79\%$ for the matched-filter
confirmation. Considering only the ``robust'' $I$-band candidates this
rate becomes $\sim75\%$ for the color slicing and $\sim85\%$ for the
matched-filter technique. Moreover, about 57\% of the candidates are
confirmed by these independent methods. This rate increases to 66\%
when only ``robust'' detections are considered. These results enable us
to adequately rank the candidate clusters for follow-up
observations. For candidates at higher redshifts ($z\geq0.7$), nearly
all ``poor'' detections, it is occasionally possible to tentatively
confirm their existence by matching $VI$-detections ($\sim26\%$).

For the confirmed clusters we find good agreement between the
redshifts estimated by the different methods with a scatter comparable
to that expected for redshifts estimated by the matched-filter
procedure. We do find that the matched-filter estimated redshifts may
be biased high by $\Delta z\sim0.1$. Correcting the redshifts by this
amount we find a very good match between the mean color of early-type
galaxies at different redshifts and those predicted by the passive
evolution models, consistent with the findings of other authors
\citep[e.g. ][]{lubin96b,stanford98,aragon-salamanca93}. These results
show that, in general the matched filter redshifts can be used as a
rough guide for a redshift selected sample.

Finally, our results are also consistent with the blind search of
clusters based on the clustering of color-selected galaxies carried
out by \citet{benoist01}.  In contrast to our analysis, which
considers the clustering of red galaxies at the position of the
matched-filter candidates, \citet{benoist01} builds a sample of cluster
candidates from the clustering of galaxies split according to
the expected colors of early-type galaxies at different redshifts. The
analysis considers wide-angular regions and the algorithm used in
identifying significant peaks is also different. Nevertheless, there
is a good agreement between the cluster sample derived and those
identified by the matched-filter.

\section{Summary}
\label{sec:conclusions}

This paper presents the first results of a systematic follow-up
program using direct imaging data to confirm EIS galaxy clusters. We
have used publicly available $V$- and $I$-band data over
$\sim8$~square degrees to investigate the reliability of the original
cluster candidates. To this end we have used two largely independent
methods based on: a direct comparison between $V$- and
$I$-matched-filter detections; second, a search for early-type cluster
members at the position of the $I$-detections using the available
$(V-I)$ color galaxy catalogs.  In the redshift interval $z\leq0.6$,
for which the $V$-band filter is most suitable, both methods provide
supporting evidence that most clusters ($\sim95$~systems or $\sim80\%$
of the sample), originally identified from the analysis of the
$I$-band galaxy catalogs, are real. Interpreting these results as a
true estimate of the expected yield of real clusters, we speculate
that when all the available $V$-band data are fully reduced and
analyzed a sample comprising $\gsim150$ clusters with $z\leq0.6$ will
be available in the southern hemisphere for statistical studies. We
also find that in this redshift range the estimated redshifts seem to
be robust within $\Delta z\sim0.1$ and can be used to select clusters
in different redshift bins.  To extend the present analysis to higher
redshifts it is necessary to use redder passbands. For this purpose we
have assembled $R$- and $JK_s$-band data for $\sim110$ and $\sim40$
candidates, respectively.

The results obtained in this paper demonstrate the usefulness of using
multi-band imaging data for a preliminary pruning of the cluster
candidates and for the selection of individual galaxies for follow-up
spectroscopic observations. As mentioned above we are currently
complementing the $V$- and $I$-data with $BRJK_s$-observations which
will allow us to carry out not only the color analysis as presented
here but also to estimate photometric redshifts. This will
significantly improve our ability to confirm clusters and to select
spectroscopic targets. Furthermore, ongoing spectroscopic observations
of EIS clusters selected in different redshift ranges will provide key
information either to confirm the results presented here or to refine
our selection methods, resolving some of the issues raised by the
present analysis.

\acknowledgements

We would like to thank the EIS Team for the great effort they have put
in producing the publicly available object catalogs from EIS and Pilot
Survey. We thank T. Beers for kindly providing his adaptive kernel
smoothing program. LFO thanks the SARC and Carlsberg Foundations for
financial support during the project period.

\bibliographystyle{apj}
\bibliography{/home/lisbeth/tex/lisbeth_ref}

\end{document}